\documentclass[conference]{IEEEtran}
\usepackage{eurosym}
\usepackage{cite}
\usepackage{graphicx}
\usepackage{epstopdf}
\usepackage{amsmath,amsfonts,amsthm,amssymb}

\ifCLASSINFOpdf
\else
\fi
\begin{document}

\title{A Generalized Labeled Multi-Bernoulli Filter for Maneuvering Targets}

\author{\IEEEauthorblockN{Yuthika Punchihewa}
\IEEEauthorblockA{School of Electrical and\\Computer Engineering\\
Curtin University of Technology\\
WA, Australia\\
Email: yuthikasgp@yahoo.com}
\and
\IEEEauthorblockN{Ba-Ngu Vo}
\IEEEauthorblockA{School of Electrical and\\Computer Engineering\\
Curtin University of Technology\\
WA, Australia\\
Email: ba-ngu.vo@curtin.edu.au}
\and
\IEEEauthorblockN{Ba-Tuong Vo}
\IEEEauthorblockA{School of Electrical and\\Computer Engineering\\
Curtin University of Technology\\
WA, Australia\\
Email: ba-tuong.vo@curtin.edu.au
}}

\maketitle

\begin{abstract}
A multiple maneuvering target system can be viewed as a Jump Markov System
(JMS) in the sense that the target movement can be modeled using different
motion models where the transition between the motion models by a particular
target follows a Markov chain probability rule. This paper describes a
Generalized Labelled Multi-Bernoulli (GLMB) filter for tracking maneuvering
targets whose movement can be modeled via such a JMS. The proposed filter is
validated with two linear and non-linear maneuvering target tracking
examples.
\end{abstract}

\IEEEpeerreviewmaketitle

\section{Introduction}

Multiple target tracking is the problem of estimating an unknown and time
varying number of trajectories from observed data. There are two main
challenges in this problem. The first is the time-varying number of targets
due to the appearance of new targets and deaths of existing targets, while
the second is the unknown association between measurements and targets,
which is further confounded by false measurements and missed detections of
actual targets \cite{YRK01,YPX11, Mahler1, Mahler14, Blackman,Barshalom}.

The Bayes optimal approach to the multi-target tracking problem is the Bayes
multi-target filter that recursively propagates the multi-target posterior
density forward in time \cite{Mahler1} incorporating both the uncertainty in
the number of objects as well as their states. Under the standard
multi-target system model (which takes into account target
births,deaths,survivals and detections,misdetections and clutter), the
multi-target posterior densities at each time are Generalized Labeled
Multi-Bernoulli (GLMB) densities \cite{GLMB1}. The $\delta $-GLMB filter 
\cite{GLMB2,HVV15_1, HVV15_2} is an analytic solution to the multi-target
Bayes filter.

While a non-maneuvering target motion can be described by a fixed model, a
combination of motion models that characterise different maneuvers may be
needed to describe the motion of a maneuvering target.Tracking a maneuvering
target in clutter is a challenging problem and is the subject of numerous
works \cite{Kiruba00,Doucet,Verca1,YRK01, YPX11, LiVSMM00, LiJilkovMM05, IMM,Pasha,Dunne13, IMM_CBMEMBER, Reuter15}. Tracking multiple maneuvering targets involves jointly
estimating the number of targets and their states at each time step in the
presence of noise, clutter, uncertainties in target maneuvers, data
association and detection. As such, this problem is extremely challenging in
both theory and implementation.

The jump Markov system (JMS) or multiple models approach has proven to be an
effective tool for single maneuvering target tracking \cite{Doucet,Verca1,Pasha,Dunne13, IMM_CBMEMBER, Reuter15}. In this
approach, the target can switch between a set of models in a Markovian
fashion. The interacting multiple model(IMM) and variable-structure IMM
(VS-IMM) estimators \cite{YRK01, YPX11, LiVSMM00, LiJilkovMM05, IMM} are two
well known single-target filtering algorithms for maneuvering targets. The
number of modes in the IMM is kept fixed, whereas in the VS-IMM the number
of modes are adaptively selected from a fixed set of modes for improved
estimation accuracy and computational efficiency.

A Probability Hypothesis Density (PHD) filter \cite{MahlerPHD2} for
maneuvering target tracking was derived in \cite{Pasha} together with a
Gaussian mixture implementation and particle implementation. As shown by
Mahler in \cite{MahlerJMS12}, this was the only mathematically valid filter
amongst severval PHD (and Cardinalized PHD) filters proposed for jump Markov
systems (JMSs) \cite{Punith08},\cite{Georgescu12}. Recently, multi-Bernoulli
and labeled multi-Bernoulli \cite{VVC09}, \cite{Reuter14} filters were also
derived for JMSs in \cite{Dunne13, IMM_CBMEMBER, Reuter15}. These filters,
however, are only approximate solutions to the Bayes multi-target filter for
maneuvering targets, and at present there are no exact solutions in the
literature.

In this paper, we propose an analytic solution to the Bayes multi-target
filter for maneuvering target tracking using JMSs. Specifically, we extend
the GLMB filter to JMSs that can be implemented via Gaussian mixture or
sequential Monte Carlo methods. In addition to being an analytic solution
and hence more accurate than approximations, the proposed solution outputs
tracks or trajectories of the targets, whereas the PHD and (unlabeled)
multi-Bernoulli filters do not. The proposed technique is verified via
numerical examples.

\section{Background}

We review JMS and the Bayes multi-target tracking filter in this section.

\subsection{JMS model for maneuvering targets}

A JMS consists of a set of parameterised state space models, whose
parameters evolve with time according to a finite state Markov chain. An
example of a maneuvering target scenario which can be successfully
represented using a JMS model is the dynamics of an aircraft, which can fly
with a nearly constant velocity motion, accelerated/decelerated motion, and
coordinated turn \cite{YRK01, YPX11}. Under a JMS framework for such a
system a target that is moving under a certain motion model at any time step
are assumed to follow the same motion model with a certain probability or
switch to a different motion model (that belongs to a set of pre-selected
motion models) with a certain probability in the next time step.

A Markovian transition probability matrix describes the probabilities with
which a particular target changes/retains the motion model in the next time
step given the motion model at current time step. Let $\vartheta
(r|r^{\prime })$ denote the probability of switching to motion model $r$
from $r^{\prime }$ as given by this markovian transition matrix, in which
the sum of the conditional probabilities of all possible motion models in
the next time step given the current model adds upto 1, i.e.,%
\begin{equation}
\sum_{r\in \mathcal{R}}\vartheta (r|r^{\prime })=1
\end{equation}%
where $R$ is the (discrete) set of motion models in the system.

Suppose that model $r$ is in effect at time $k$, then the state transition
density from $\zeta ^{\prime }$, at time $k-1$, to $\zeta $, at time $k$, is
denoted by $\phi _{k|k-1}(\zeta |\zeta ^{\prime },r)$, and the likelihood of 
$\zeta $ generating the measurement $z$ is denoted by $\gamma _{k}(z|\zeta
,r)$ \cite{YRK01, YPX11, Ristic04}. Moreover, the joint transition of the
state and the motion model assumes the form:%
\begin{equation}
f_{k|k-1}(\zeta ,r|\zeta ^{\prime },r^{\prime })=\phi _{k|k-1}(\zeta |\zeta
^{\prime },r)\vartheta (r|r^{\prime }).
\end{equation}%
In general, the measurement can also depend on the model $r$ and hence the
likelihood function becomes $g_{k}(z|\zeta ,r)$. Note that by defining the
augmented system state as $x=(\zeta ,r\mathbf{)}$ a JMS model can be written
as a standard state space model.

JMS models are not only useful for tracking maneuvering targets, but are
also useful in the estimation of unknown clutter parameters \cite{MahlerclutterCPHD11,RobustMB13}.

\subsection{Bayes multi-target tracking filter}

In the Bayes multi-target tracking filter, the state of a target includes an
ordered pair of integers $\ell =(k,i)$, where $k$ is the time of birth, and $%
i$ is a unique index to distinguish targets born at the same time. The label
space for targets born at time $k$ is denoted as $\mathbb{L}_{k}$ and the
label space for targets at time $k$ (including those born prior to $k$) is
denoted as $\mathbb{L}_{0:k}$. Note that $\mathbb{L}_{0:k}=\mathbb{L}%
_{0:k-1}\cup \mathbb{L}_{k}$, and that $\mathbb{L}_{0:k-1}$ and $\mathbb{L}%
_{k}$ are disjoint. An existing target at time $k$ has state $\mathbf{x}%
=(x,\ell )$ consisting of the kinematic/feature $x$ and label $\ell \in 
\mathbb{L}_{0:k}$. A multi-target state $\mathbf{X}$ (uppercase notation) is
a finite set of single-target states.

All information about the multi-target state at time $k$ is contained in $%
\mathbf{\pi }_{k}$, the posterior density of the multi-target state
conditioned on $Z_{1:k}=(Z_{1},...,Z_{k})$, the measurement history upto
time $k$, where $Z_{k}$ is the finite set of measurements received at time $%
k $. The Bayes multi-target tracking filter consists of a prediction step %
\eqref{eq1} and an update step \eqref{eq2}, which propagate the multi-target
posterior/filtering density forward in time. Note that the integral in this
case is the \emph{set integral} from finite set statistics \cite{Mahler1}. 
\begin{equation}
\mathbf{\pi }_{k|k-1}(\mathbf{X}_{k})\!=\!\!\int \!\mathbf{f}_{k|k-1}(%
\mathbf{X}_{k}|\mathbf{X})\mathbf{\pi }_{k-1}(\mathbf{X})\delta \mathbf{X}
\label{eq1}
\end{equation}%
\begin{equation}
\mathbf{\pi }_{k}(\mathbf{X}_{k})=\frac{g_{k}(Z_{k}|\mathbf{X}_{k})\mathbf{%
\pi }_{k|k-1}(\mathbf{X}_{k})}{\int g_{k}(Z_{k}|\mathbf{X})\mathbf{\pi }%
_{k|k-1}(\mathbf{X})\delta \mathbf{X}}  \label{eq2}
\end{equation}%
where $\mathbf{f}_{k|k-1}(\cdot |\cdot )$ denotes the multi-target
transition kernel from time $k-1$\textit{\ }to $k$, and $g_{k}(\cdot |\cdot
) $ denotes the likelihood function at time $k$. Note that for compactness
we omitted dependence on the measurement history from $\mathbf{\pi }_{k|k-1}$
and $\mathbf{\pi }_{k}$. Note that the same multi-target recursion %
\eqref{eq1}-\eqref{eq2} also holds for multi-target states without labels.

A generic particle implementation of the multi-target Bayes recursions %
\eqref{eq1}-\eqref{eq2} (for both labeled and unlabeled multi-target states)
was given in \cite{VoAES}, while analytic approximations for unlabeled
multi-target states, such as the PHD, Cardinalized PHD and multi-Bernoulli
filters were proposed in \cite%
{Mahler1,MahlerCPHDAES,VoMaGMPHD05,VoGaussianCPHD07,VVC09,VVPS10}. The GLMB
filter \cite{GLMB1}, \cite{GLMB2} is an analytic solution to the
multi-target Bayes recursions \eqref{eq1}-\eqref{eq2}.

\section{JMS-GLMB filtering}

We start this subsection with some notations. For the labels of a
multi-target state $\mathbf{X}$ to be distinct, we require $\mathbf{X}$ and
the set of labels of $\mathbf{X}$, denoted as $\mathcal{L}(\mathbf{X})$, to
have the same cardinality, .i.e. the same number of elements. Hence, we
define the \emph{distinct label indicator} as the function%
\begin{equation*}
\Delta (\mathbf{X})\triangleq \delta _{|\mathbf{X}|}[|\mathcal{L(}\mathbf{X}%
)|],
\end{equation*}%
where $|Y|$ denotes the cardinality of the set $Y$, and $\delta _{n}[m]$
denotes the Kronecker delta. The \emph{indicator function} is defined as as 
\begin{equation*}
1_{Y}(x)\triangleq \Big\{{}%
\begin{array}{ll}
1, & \text{if }x{\in Y} \\ 
0, & \text{otherwise }{}%
\end{array}%
.
\end{equation*}%
For any finite set $Y$, and test function $h\leq 1$, the multi-object
exponential is defined by%
\begin{equation*}
h^{Y}\triangleq \prod_{y\in Y}h(y),
\end{equation*}%
with $h^{\emptyset }$ $=1$ by convention. We also use the standard inner
production notation%
\begin{equation*}
\left\langle f,g\right\rangle =\int f(x)g(x)dx,
\end{equation*}%
for any real functions $f$ and $g$.

An \emph{association map} at time $k$ is a function $\theta :\mathbb{L}%
_{0:k}\rightarrow \{0,1,...,|Z_{k}|\}$ such that $\theta (\ell )=\theta
(\ell ^{\prime })>0~$implies$~\ell =\ell ^{\prime }$. Such a function can be
regarded as an assignment of labels to measurements, with undetected labels
assigned to $0$. The set\ of all such association maps is denoted as $\Theta
_{k}$; the subset of association maps with domain $L$ is denoted by $\Theta
_{k}(L)$; and $\Theta _{0:k}\triangleq \Theta _{0}\times ...\times \Theta
_{k}$ denotes the space of association map history.

\subsection{GLMB filter}

In the GLMB filter, the multi-target filtering density at time $k-1$ is a
GLMB of the form: 
\begin{equation}
\mathbf{\pi }_{k-1}(\mathbf{X})=\Delta (\mathbf{X})\!\!\!\sum_{\xi \in
\Theta _{0:k\!-\!1}}\!\!\!\!w_{k-1}^{(\xi )}(\mathcal{L(}\mathbf{X}%
))[p_{k-1}^{(\xi )}]^{\mathbf{X}},  \label{eq:GLMB_prev}
\end{equation}%
where each $\xi =(\theta _{0},...,\theta _{k-1})\in \Theta _{0:k-1}$
represents a history of association maps up to time $k-1$; each weight $%
w_{k-1}^{(\xi )}(L)$ is non-negative with 
\begin{equation*}
\sum_{L\subseteq \mathbb{L}_{0:k-1}}\sum_{\xi \in \Theta
_{0:k-1}}w_{k-1}^{(\xi )}(L)=1,
\end{equation*}%
and each $p_{k-1}^{(\xi )}(\cdot ,\ell )$ is a probability density.

Given a GLMB filtering density, a tractable suboptimal multi-target estimate
is obtained by the following proceedure: determine the maximum a posteriori
cardinality estimate $n^{\ast }$ from the cardinality distribution%
\begin{equation}
\rho _{k-1}(n)=\sum_{L\subseteq \mathbb{L}_{0:k-1}}\sum_{\xi \in \Theta
_{0:k-1}}\delta _{n}[|L|]w_{k-1}^{(\xi )}(L);  \label{eq:GLMBCard}
\end{equation}%
determine the label set $L^{\ast }$ and $\xi ^{\ast }$ with highest weight $%
w_{k-1}^{(\xi ^{\ast })}(L^{\ast })$ among those with cardinality $n^{\ast }$%
; determine the expected values of the states from $p_{k-1}^{(\xi ^{\ast
})}(\cdot ,\ell )$, $\ell \in L^{\ast }$ \cite{GLMB1}.

The GLMB density is a conjugate prior with respect to the standard
multi-target likelihood function and is also closed under the multi-target
prediction \cite{GLMB1}. Under the standard multi-target transition model,
if the multi-target filtering density, at the previous time, $\mathbf{\pi }%
_{k-1}$ is a GLMB of the form (\ref{eq:GLMB_prev}), then the multi-target
prediction density $\mathbf{\pi }_{k|k-1}$ is a GLMB of the form (\ref%
{eq:GLMBpred}) given by \cite{GLMB1}.%
\begin{equation}
\mathbf{\pi }_{k|k-1}(\mathbf{X})=\Delta (\mathbf{X})\!\!\!\sum_{\xi \in
\Theta _{0:k-1}}\!\!w_{k|k-1}^{(\xi )}(\mathcal{L(}\mathbf{X}%
))[p_{k|k-1}^{(\xi )}]^{\mathbf{X}},  \label{eq:GLMBpred}
\end{equation}%
where\allowdisplaybreaks%
\begin{eqnarray*}
\!\!\!\!\!\!\!\!\!\!w_{k|k\!-\!1\!}^{(\xi )}(L)\!\!\!\!
&=&\!\!\!\!w_{S,k|k-1}^{(\xi )}(L\cap \mathbb{L}_{0:k-1})w_{B,k}(L\cap 
\mathbb{L}_{k}), \\
\!\!\!\!\!\!\!\!\!\!p_{k|k\!-\!1\!}^{(\xi )}(x,\ell )\!\!\!\! &=&\!\!\!\!1_{%
\mathbb{L}_{0:k\!-\!1}\!}(\ell )p_{S,k|k\!-\!1\!}^{(\xi )\!}(x,\ell )\!+\!1_{%
\mathbb{L}_{k}\!}(\ell )p_{B,k}(x,\ell ), \\
\!\!\!\!\!\!\!\!\!\!w_{S,k|k\!-\!1\!}^{(\xi )}(L)\!\!\!\! &=&\!\!\!\![\bar{P}%
_{S,k|k\!-\!1}^{(\xi )}]^{L}\!\sum_{I\supseteq L}[1\!-\!\bar{P}%
_{S,k|k\!-\!1}^{(\xi )}]^{I\!-\!L}w_{k\!-\!1\!}^{(\xi )}(I), \\
\!\!\!\!\!\!\!\!\!\!\bar{P}_{S,k|k\!-\!1\!}^{(\xi )}(\ell )\!\!\!\!
&=&\!\!\!\!\left\langle P_{S,k|k-1}(\cdot ,\ell ),p_{k-1}^{(\xi )}(\cdot
,\ell )\right\rangle , \\
\!\!\!\!\!\!\!\!\!\!p_{S,k|k\!-\!1\!}^{(\xi )}(x,\ell )\!\!\!\! &=&\!\!\!\!%
\frac{\left\langle P_{S,k|k\!-\!1}(\cdot ,\ell )f_{k|k\!-\!1\!}(x|\cdot
,\ell ),p_{k\!-\!1}^{(\xi )}(\cdot ,\ell )\right\rangle }{\bar{P}%
_{S,k|k-1}^{(\xi )}(\ell )}, \\
\!\!\!\!\!\!\!\!\!\!P_{S,k|k-1}(x,\ell )\!\!\!\! &=&\!\!\!\!\text{%
probability of survival to time }k\text{ of a target } \\
&&\!\!\!\!\text{with previous state }(x,\ell ), \\
f_{k|k\!-\!1\!}(x|x^{\prime },\ell )\!\!\!\! &=&\!\!\!\!\text{transition
density of feature }x^{\prime }\text{ at time } \\
&&\!\!\!\!k-1\text{ to }x\text{ at time }k\text{ for target with label }\ell 
\text{, } \\
\!\!\!\!\!\!\!\!\!\!w_{B,k}(L)\!\!\!\! &=&\!\!\!\!\text{probability of
targets with labels }L\text{ being } \\
&&\!\!\!\!\text{born at time }k\text{,} \\
\!\!\!\!\!\!\!\!\!\!p_{B,k}(x,\ell )\!\!\!\! &=&\!\!\!\!\text{probability
density of the feature }x\text{ of a} \\
&&\!\!\!\!\text{new target born at time }k\text{ with label }\ell \text{.}
\end{eqnarray*}

Moreover, under the standard multi-target measurement model, the
multi-target filtering density $\mathbf{\pi }_{k}$ is a GLMB given by%
\begin{equation}
\mathbf{\pi }_{k}\!(\mathbf{X})=\Delta \!(\mathbf{X})\!\!\!\!\!\!\sum_{\xi
\in \Theta _{0:k\!-\!1}}\sum\limits_{\theta \in \Theta _{k}}\!w_{k}^{\!(\xi
,\theta )\!}(\mathcal{L(}\mathbf{X})|Z_{k})[p^{\!(\xi ,\theta )\!}(\cdot
|Z_{k})]^{\mathbf{X}}\!\!,  \label{eq:GLMBupdate}
\end{equation}%
where \allowdisplaybreaks%
\begin{eqnarray*}
w_{k}^{(\xi ,\theta )\!}(L|Z)\!\!\! &\propto &\!\!\!1_{\Theta
_{k}\!(L)}(\theta )[\bar{\Psi}_{Z,k}^{(\xi ,\theta )}]^{L}w_{k|k-1}^{(\xi
)}(L), \\
p_{k}^{\!(\xi ,\theta )\!}(x,\ell |Z)\!\!\! &=&\!\!\!\frac{\Psi
_{Z,k}^{(\theta )}(x,\ell )p_{k|k-1}^{(\xi )}(x,\ell )}{\bar{\Psi}%
_{Z,k}^{(\xi ,\theta )}(\ell )} \\
\bar{\Psi}_{Z,k}^{(\xi ,\theta )}(\ell )\!\!\! &=&\!\!\!\left\langle \Psi
_{Z,k}^{(\theta )}(\cdot ,\ell ),p_{k|k-1}^{(\xi )}(\cdot ,\ell
)\right\rangle , \\
\Psi _{\{z_{1},...,z_{m}\},k}^{(\theta )}(x,\ell )\!\!\! &=&\!\!\!\left\{ 
\begin{array}{ll}
\frac{P_{D,k}(x,\ell )g_{k}(z_{\theta (\ell )}|x,\ell )}{\kappa
_{k}(z_{\theta (\ell )})}, & \text{if }{\small \theta (\ell )>0} \\ 
{\small 1-P}_{D,k}{\small (}x{\small ,\ell )}, & \text{if }{\small \theta
(\ell )=0}%
\end{array}%
\right. \\
P_{D,k}(x,\ell )\!\!\! &=&\!\!\!\text{probability of detection at time }k \\
&&\!\!\!\text{of a target with state }(x,\ell ), \\
g_{k}(z|x,\ell )\!\!\! &=&\!\!\!\text{likelihood that at time }k\text{
target with} \\
&&\!\!\!\text{state }(x,\ell )\text{ generate measurement }z, \\
\kappa _{k}\!\!\! &=&\!\!\!\text{intensity function of Poisson clutter} \\
&&\!\!\!\text{at time }k\text{ }
\end{eqnarray*}

The GLMB recursion above is the first analytic solution to the Bayes
multitarget filter. Truncating the GLMB sum is needed to manage the growing
the number of components in the GLMB filter \cite{GLMB2}.

\subsection{GLMB filter for Manuevering Targets}

We define the (labeled) state of a manuevering target to include the
kinematic/feature $\zeta $, the motion model index $r$, and the label $\ell $%
, i.e., $\mathbf{x}=(\zeta ,r,\ell )$, which can be modeled as a JMS. Note
that the label of each target remains constant throughout it's life even
though it is part of the state vector. Hence the JMS state equations for a
target with label $\ell $ are indexed by $\ell $, i.e., $\phi
_{k|k-1}^{(\ell )}(\zeta |\zeta ^{\prime },r)$ and\ $\gamma _{k}^{(\ell
)}(z|\zeta ,r)$. The new state of a surviving target will also be governed
by the probability of the target transitioning to that motion model from the
previous model in addition to the probability of survival and the relevant
state transtition function. Consequently, the joint transition and
likelihood function for the state and the model index are given by, 
\begin{eqnarray}
f_{k|k-1}(\zeta ,r|\zeta ^{\prime },r^{\prime },\ell ) &=&\phi
_{k|k-1}^{(\ell )}(\zeta |\zeta ^{\prime },r)\times \vartheta (r|r^{\prime })
\label{eq:JMS-GLMB} \\
g_{k}(z|\zeta ,r,\ell ) &=&\gamma _{k}^{(\ell )}(z|\zeta ,r)
\label{eq:JMS-GLMB1}
\end{eqnarray}

Substituting (\ref{eq:JMS-GLMB}) and (\ref{eq:JMS-GLMB1}) into the GLMB
prediction and update equations yields the GLMB filter for maneuvering
targets. Note that since $x=(\zeta ,r)$ 
\begin{equation*}
\int f(x)dx=\sum_{r\in \mathcal{R}}\int f(\zeta ,r)d\zeta .
\end{equation*}

The state extraction is akin to the single model system. To estimate the
motion model for each label, we select the motion model that maximizes the
marginal probability of that model over the entire density for that label,
i.e., for label $\ell $ of component $\xi $, the estimated motion model $%
\hat{r}$ is given by \eqref{eq8}. 
\begin{equation}
\hat{r}=\underset{r}{\mbox{argmax}}\;\int p^{(\xi )}( \zeta,r,\ell )d\zeta
\label{eq8}
\end{equation}

\subsection{Analytic Solution}

Consider the special case where the target birth model, motion models and
observation model are all linear models with Gaussian noise. Given that the
posterior density at time $k-1$ is of the form (\ref{eq:GLMB_prev}) with $%
\mathbf{x}=(\zeta ,r,\ell )$, the GLMB filter prediction equation can be
explicitly written as%
\begin{equation}
\mathbf{\pi }_{k|k-1}(\mathbf{X})=\Delta (\mathbf{X})\!\!\!\sum_{\xi \in
\Theta _{0:k-1}}\!\!w_{k|k-1}^{(\xi )}(\mathcal{L(}\mathbf{X}%
))[p_{k|k-1}^{(\xi )}]^{\mathbf{X}},  \label{eq:GLMBpred}
\end{equation}%
where\allowdisplaybreaks%
\begin{eqnarray*}
\!\!\!\!\!\!\!\!\!\!w_{k|k\!-\!1\!}^{(\xi )}(L)\!\!\!\!
&=&\!\!\!\!w_{S,k|k-1}^{(\xi )}(L\cap \mathbb{L}_{0:k-1})w_{B,k}(L\cap 
\mathbb{L}_{k}), \\
\!\!\!\!\!\!\!\!\!\!p_{k|k\!-\!1\!}^{(\xi )}(\zeta \!,r\!,\ell \!)\!\!\!\!
&=&\!\!\!\!1_{\mathbb{L}_{0:k\!-\!1}\!}(\ell )p_{S,k|k\!-\!1\!}^{(\xi
)\!}(\zeta \!,r\!,\ell \!)\!\!+\!\!1_{\mathbb{L}_{k}\!}(\ell )p_{B,k}(\zeta
\!,r\!,\ell \!), \\
\!\!\!\!\!\!\!\!\!\!w_{S,k|k\!-\!1\!}^{(\xi )}(L)\!\!\!\! &=&\!\!\!\![\bar{P}%
_{S,k|k\!-\!1}^{(\xi )}]^{L}\!\sum_{I\supseteq L}[1\!-\!\bar{P}%
_{S,k|k\!-\!1}^{(\xi )}]^{I\!-\!L}w_{k\!-\!1\!}^{(\xi )}(I), \\
\!\!\!\!\!\!\!\!\!\!\bar{P}_{S,k|k\!-\!1\!}^{(\xi )}(\ell )\!\!\!\!
&=&\!\!\!\!\sum_{r\in R}\bar{P}_{S,k|k\!-\!1\!}^{(\xi )}(r,\ell ), \\
\!\!\!\!\!\!\!\!\!\!\bar{P}_{S,k|k\!-\!1\!}^{(\xi )}(r,\ell )\!\!\!\!
&=&\!\!\!\!\left\langle P_{S,k|k-1}(\cdot ,r,\ell ),p_{k-1}^{(\xi )}(\cdot
,r,\ell )\right\rangle , \\
\!\!\!\!\!\!\!\!\!\!p_{S,k|k\!-\!1\!}^{(\xi )}(\zeta ,r,\ell )\!\!\!\!
&=&\!\!\!\!\!\!\frac{\underset{r^{\prime }\in R}{\sum }\!\!\!\left\langle
\!\!P_{S,k|k\!-\!1}(\cdot ,\ell )f_{k|k\!-\!1\!}(\!\zeta ,r|\!\cdotp\!,r^{\prime
}\!,\ell ),p_{k\!-\!1}^{(\xi )}(\!\cdot\!,\!\ell )\!\!\right\rangle \!\!}{\bar{P}%
_{S,k|k-1}^{(\xi )}(\ell )}, \\
\!\!\!\!\!\!\!\!\!\!P_{S,k|k-1}(\zeta ,r,\ell )\!\!\!\! &=&\!\!\!\!\text{%
probability of survival to time }k\text{ of a } \\
&&\!\!\!\!\text{target with previous labeled state }(\zeta ,r,\ell ), \\
f_{k|k\!-\!1\!}(\zeta ,r|\zeta ^{\prime },r^{\prime },\ell )\!\!\!\!
&=&\!\!\!\mathcal{N}(\zeta ;F^{(r)}\zeta ^{\prime },Q_{F}^{(r)})\times
\vartheta (r|r^{\prime }) \\
F^{(r)}\!\!\! &=&\!\!\!\text{state transition matrix of motion mode\!l }r, \\
Q_{F}^{(r)}\!\!\! &=&\!\!\!\text{covariance matrix of motion model }r, \\
\!\!\!\!\!\!\!\!\!\!w_{B,k}(L)\!\!\!\! &=&\!\!\!\!\text{probability of
targets with labels }L\text{ being } \\
&&\!\!\!\!\text{born at time }k, \\
\!\!\!\!\!\!\!\!\!\!p_{B,k}(\zeta ,r,\ell )\!\!\!\! &=&\!\!\!\!\mathcal{N}%
(\zeta ;m^{(i)},Q_{B}^{(i)})\times \vartheta ^{(i)}(r) \\
\vartheta ^{(i)}(r) &=&\!\!\text{probability that a target born at birth} \\
&&\!\!\!\!\text{ region }i\text{ possesses motion model }r, \\
m^{(i)}\!\!\! &=&\!\!\!\text{mean of birth region }i, \\
Q_{B}^{(i)}\!\!\! &=&\!\!\!\text{covariance of birth region }i,
\end{eqnarray*}

Moreover, GLMB update formula can be written explicitly as%
\begin{equation}
\mathbf{\pi }_{k}\!(\mathbf{X})=\Delta \!(\mathbf{X})\!\!\!\!\!\!\sum_{\xi
\in \Theta _{0:k\!-\!1}}\sum\limits_{\theta \in \Theta _{k}}\!w_{k}^{\!(\xi
,\theta )\!}(\mathcal{L(}\mathbf{X})|Z_{k})[p^{\!(\xi ,\theta )\!}(\cdot
|Z_{k})]^{\mathbf{X}}\!\!,  \label{eq:GLMBupdate}
\end{equation}%
where \allowdisplaybreaks%
\begin{eqnarray*}
w_{k}^{(\xi ,\theta )\!}(L|Z)\!\!\! &\propto &\!\!\!1_{\Theta
_{k}\!(L)}(\theta )[\bar{\Psi}_{Z,k}^{(\xi ,\theta )}]^{L}w_{k|k-1}^{(\xi
)}(L), \\
p_{k}^{\!(\xi ,\theta )\!}(\zeta ,r,\ell |Z)\!\!\!\! &=&\!\!\!\frac{\Psi
_{Z,k}^{(\theta )}(\zeta ,r,\ell )p_{k|k-1}^{(\xi )}(\zeta ,r,\ell )}{\bar{%
\Psi}_{Z,k}^{(\xi ,\theta )}(\ell )} \\
\bar{\Psi}_{Z,k}^{(\xi ,\theta )}(\ell )\!\!\!\! &=&\!\!\!\sum_{r\in
R}\left\langle \Psi _{Z,k}^{(\theta )}(\cdotp,r,\ell ),p_{k|k-1}^{(\xi
)}(\cdot ,r,\ell )\right\rangle , \\
\Psi _{\{z_{1},...,z_{m}\},k}^{(\theta )}(\zeta ,r,\ell )\!\!\!\!
&=&\!\!\!\!\left\{ 
\begin{array}{ll}
\!\!\!\frac{P_{D,k}(\zeta ,r,\ell )g_{k}(z_{\theta (\ell )}|\zeta ,r,\ell )}{%
\kappa _{k}(z_{\theta (\ell )})}, & \!\!\!\!\text{if }{\small \theta (\ell
)>0} \\ 
\!\!\!{\small 1-P}_{D,k}{\small (}\zeta ,r{\small ,\ell )}, & \!\!\!\!\text{%
if }{\small \theta (\ell )=0}%
\end{array}%
\right.  \\
P_{D,k}(\zeta ,r,\ell )\!\!\! &=&\!\!\!\!\text{probability of detection at
time }k \\
&&\!\!\!\text{of a target with state }(\zeta ,r,\ell ), \\
g_{k}(z|\zeta ,r,\ell )\!\!\! &=&\!\!\!\!\mathcal{N}(z;H^{(r)}\zeta
,Q_{H}^{(r)}) \\
\kappa _{k}\!\!\! &=&\!\!\!\!\text{intensity function of Poisson clutter}, \\
H^{(r)}\!\!\! &=&\!\!\!\!\text{likelihood matrix for targets } \\
&&\!\!\!\!\text{moving under motion model r}, \\
Q_{H}^{(r)}\!\!\! &=&\!\!\!\!\text{covariance matrix of likelihood for } \\
&&\!\!\!\!\text{targets \negthinspace moving \negthinspace under
\negthinspace motion \negthinspace model r}.
\end{eqnarray*}

For mildly non-linear motion models and  measurement models, the unscented
Kalman Filter (UKF) \cite{Ristic04,UKF} can been utilized for predicting and
updating each Gaussian component in the mixture forward. Alternatively,
instead of a making use of a Gaussian mixture to represent the posterior
density of each track in a hypothesis, a particle filter can be employed.
Instead of a Gaussian mixture, the density is represented using a set of
particles which are propagated forward under the different motion models
with adjusted weights for each particle. As in the case of the Gaussian
mixture, the number of particles in the density increase by threefold during
each prediction forward. Thus resampling needs to be carried out to discard
particles with negligible weights and keep the total count of particles in
control.

\subsection{Implementation Issues}

In the above solution it is evident that the posterior density for each
track is a Gaussian mixture, with each mixture component relating to one of
the motion models present. For a particular track, at each new time step the
posterior is predicted forward for all motion models present in the system,
thereby generating a new Gaussian mixture. The weight of each new component
will be the weight of the parent component multiplied by the probability of
switching to the corresponding motion model. As a result the number of
mixture components escalates exponentially. Hence extensive pruning and
merging must be carried out for each track in each GLMB hypothesis after the
update step to keep the computation managable.

\section{ Simulation Results}

In this section we demonstrate the use of the proposed JMS-GLMB solution via
two multiple manuevering target tracking examples.

\textit{Linear Example}: The kinematic state of each target in this example
consists of cartesian x and y coordinates and their respective velocities. $%
T=5s$ is the sampling interval. The observation area is a [-60, 60] $\times $
[-60, 60] $km^{2}$ area. The JMS used in the simulation consists of three
types of motion models viz. constant velocity, right turn (coordinated turn
with a $3^{\circ }$ angle), and left turn (coordinated turn with a $%
-3^{\circ }$ angle). The state transition matrices for the three models are
obtained via substituting $\omega =0$, $\omega =5\pi /180$ and $\omega
=-5\pi /180$ in equation \eqref{m2} respectively.The process noise
co-variance $Q_{L}$ is given in \eqref{procnoise} with $\sigma
_{v1}=5ms^{-1},\sigma _{v2}=\sigma _{v3}=20ms^{-1}$. The markovian motion
model switching probability matrix is given in \eqref{mvt}.

\begin{equation}  \label{m1}
F_1 = 
\begin{bmatrix}
1 & T & 0 & 0 \\ 
0 & 1 & 0 & 0 \\ 
0 & 0 & 1 & T \\ 
0 & 0 & 0 & 1%
\end{bmatrix}%
\end{equation}

\begin{equation}
F_{2}(\omega)=%
\begin{bmatrix}
1 & sin(T\omega )/\omega & 0 & (cos(T\omega )-1)/\omega \\ 
0 & cos(T\omega ) & 0 & -sin(T\omega ) \\ 
0 & -(cos(T\omega )-1)/\omega & 1 & sin(T\omega )/\omega \\ 
0 & sin(T\omega ) & 0 & cos(T\omega )%
\end{bmatrix}
\label{m2}
\end{equation}

\begin{equation}  \label{procnoise}
Q_L = \sigma_{vr}^2 
\begin{bmatrix}
T^4/4 & T^3/2 & 0 & 0 \\ 
T^3/2 & T^2 & 0 & 0 \\ 
0 & 0 & T^4/4 & T^3/2 \\ 
0 & 0 & T^3/2 & T^2%
\end{bmatrix}%
\end{equation}

\begin{equation}  \label{mvt}
\vartheta(r^{\prime }|r) = M( r, r^{\prime }) \text{ where } M = 
\begin{bmatrix}
0.8 & 0.1 & 0.1 \\ 
0.2 & 0.8 & 0 \\ 
0.2 & 0 & 0.8%
\end{bmatrix}%
\end{equation}

Targets are spontaneously born at three pre-defined Gaussian birth locations 
$\mathcal{N}(m_{1},P_{L}),\mathcal{N}(m_{2},P_{L}),\mathcal{N}(m_{3},P_{L})$
where. 
\begin{equation*}
m_{1}=[:40000,0,-50000,0],::m_{2}=[:-50000,0,40000,0]
\end{equation*}%
\begin{equation*}
m_{3}=[-10000,0,0,0],P_{L}\!=\!diag([\!1000,\!300,\!1000,\!300]).
\end{equation*}%
Targets are born from each location at each time step with a probability of
0.2 and the initial motion model is model 1.

The $x$ and $y$ corrdinates of the targets are observed by a single sensor
located at (0, 0) with probability of detection $P_{D}=0.97$ (observation
matrix H given in \eqref{obs}). The measurements are subjected to zero mean
noise with a covariance of $\sigma _{h}^{2}I_{2}$ where $\sigma _{h}=40m $
and $I_{2}$ is the identity matrix of dimestion 2. Clutter is modeled as a
uniform Poisson with an average number of 60 measurements per scan.

\begin{equation}
H=%
\begin{bmatrix}
1 & 0 & 0 & 0 \\ 
0 & 0 & 1 & 0%
\end{bmatrix}
\label{obs}
\end{equation}

\begin{figure}[tbp]
\includegraphics[scale = 0.5, trim= 0 10 50 50]{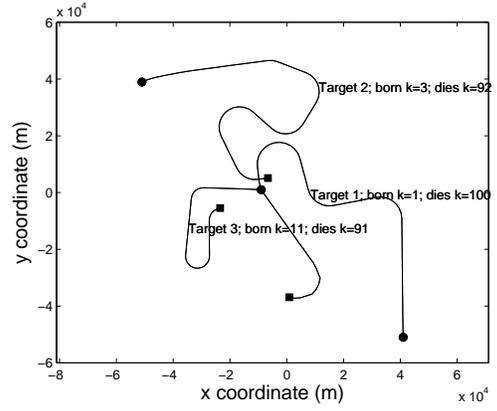}
\caption{True Target Trajectories - Linear Example}
\label{traj1}
\end{figure}

\begin{figure}[tbp]
\includegraphics[scale=0.23, trim= 120 100 100 150]{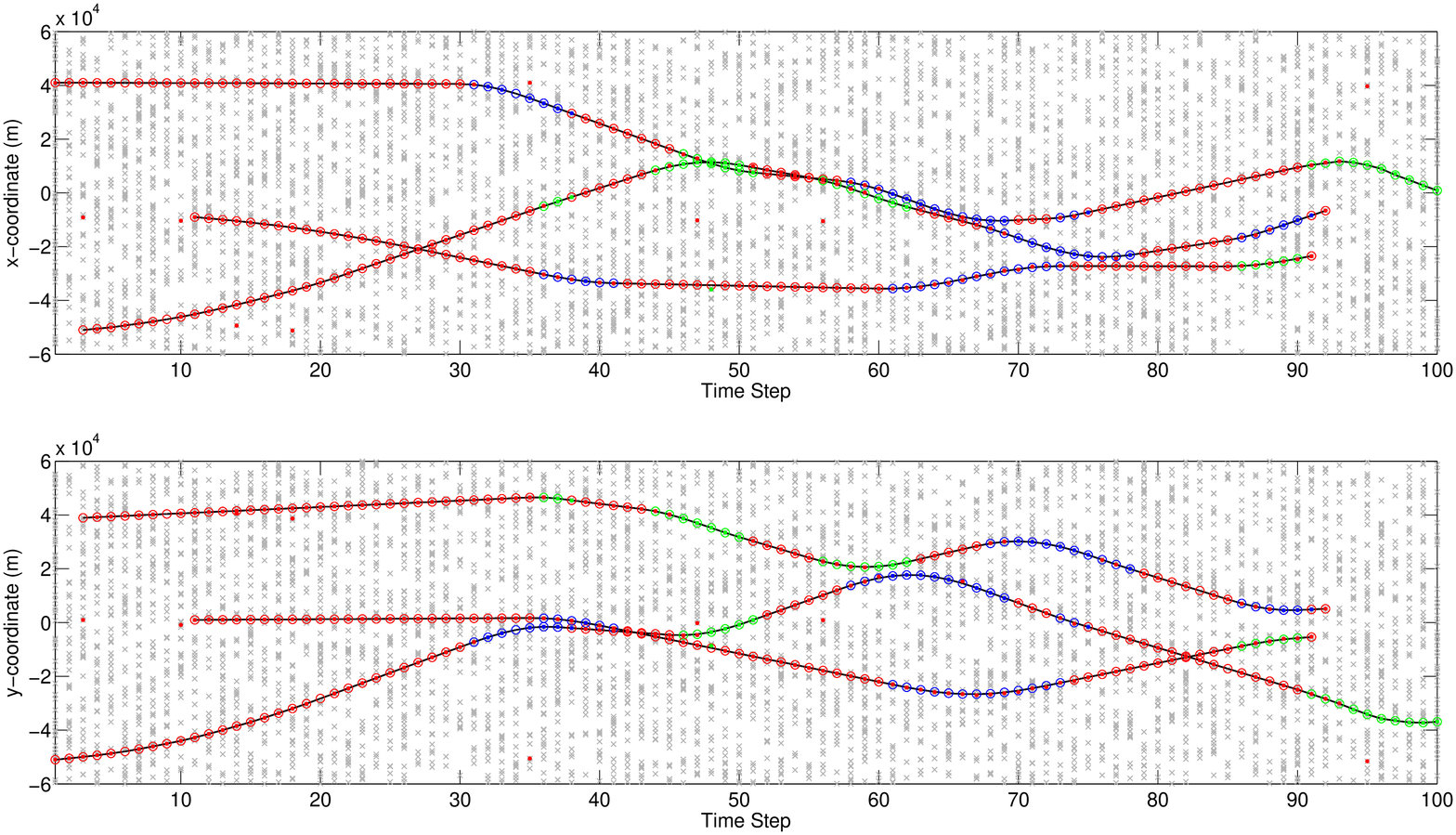}
\caption{ Position Estimates - Linear Example.}
\label{est}
\end{figure}

Figure (\ref{traj1}) shows the trajectories of three targets born at
different time steps in a simlation run. Fig.(\ref{est}) illustrates the
estimated coordinates colour coded in red (constant velocity), blue (right
turn) and green (left turn) to indicate the estimated motion models along
with the true path (coninous lines) and measurements (grey crosses).

\begin{figure}[tbp]
\includegraphics[scale=0.5, trim= 0 10 50 10]{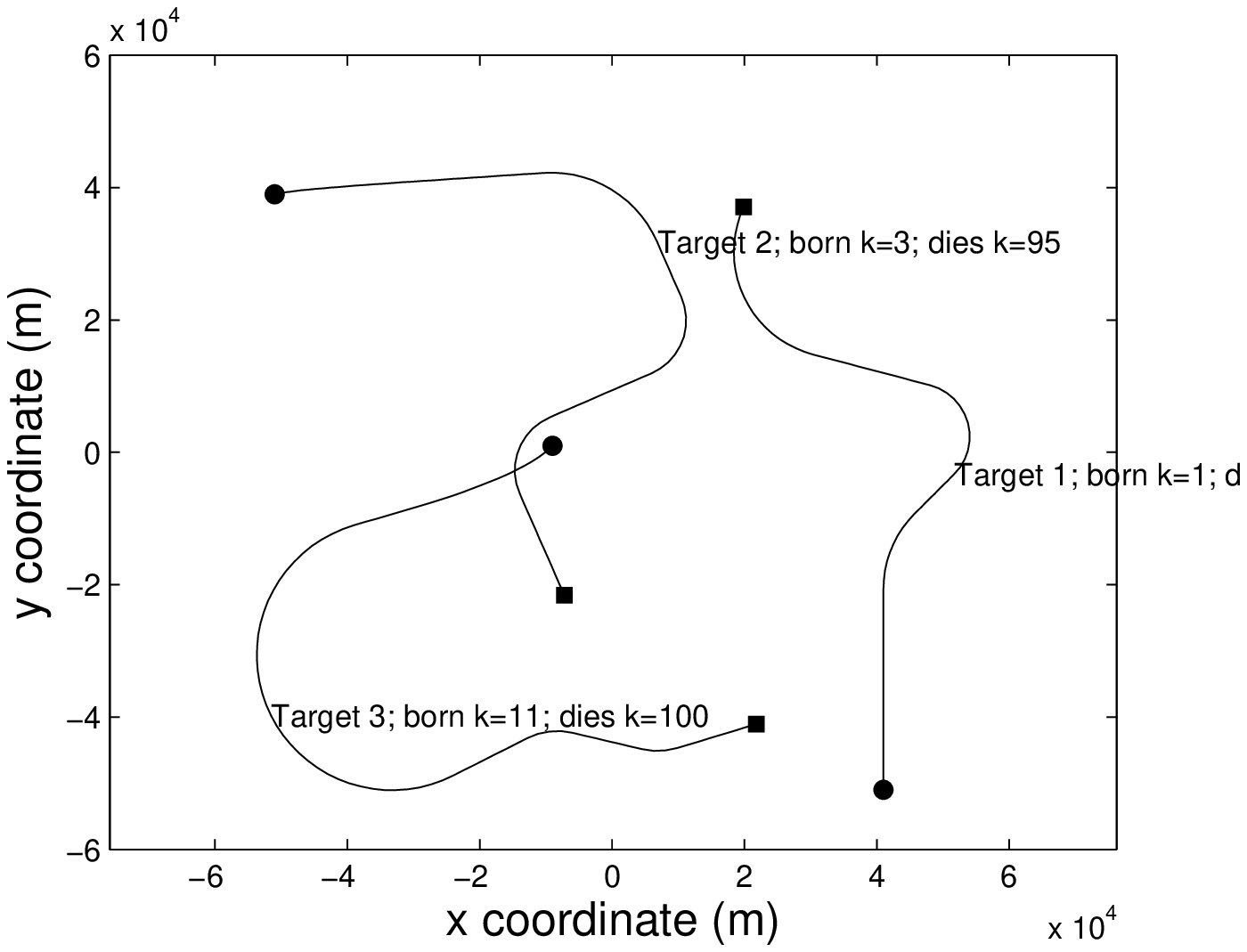}
\caption{True Target Trajectories - Non-linear Example}
\label{traj2}
\end{figure}

\begin{figure}[tbp]
\includegraphics[scale=0.4, trim= 20 0 50 20]{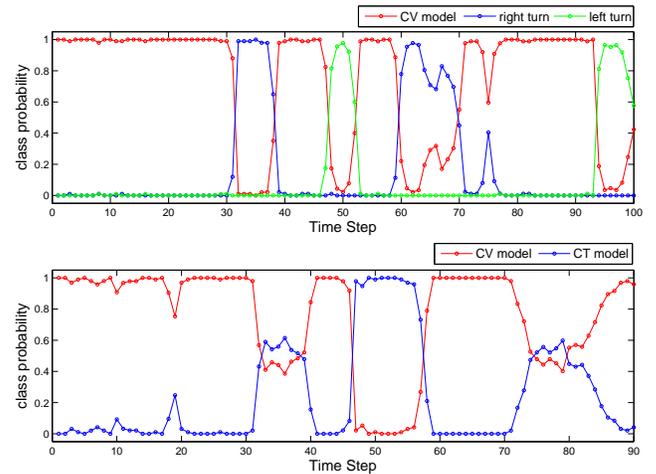}
\caption{ Probability of estimating each motion model for target 1 in linear
example (above) and non-Linear example (below) (100 mc runs).}
\label{tgt1}
\end{figure}

\begin{figure}[tbp]
\includegraphics[scale=0.38, trim= 20 20 120 20]{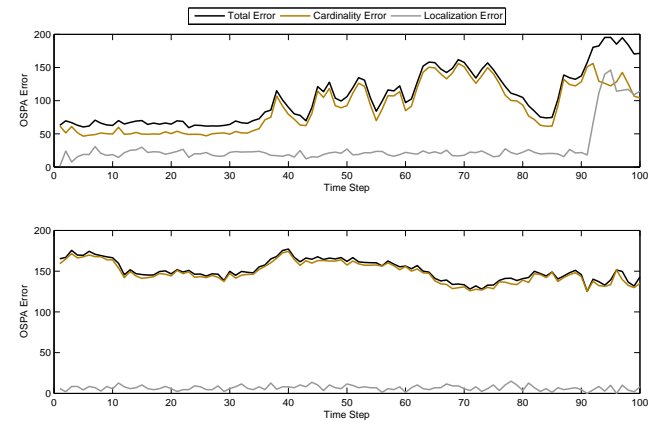}
\caption{OSPA calculation with C = 200m P = 2 for linear example(above) and
non-linear example (below)(100 mc runs).}
\label{ospa}
\end{figure}

The Optimal Subpattern Assignment Metric (OSPA)\cite{OSPApaper} values
calculated for 100 monte carlo runs for the linear example are shown in the 
top graph of fig.(\ref{ospa}). The top graph of figure (\ref{tgt1}) shows the
probabilities of estimating each motion model (colour coded) in each time
step for target 1. For example, between time steps 1 to 30, constant velocity model (red)
has a higher probability (above 0.9 in most time steps) of being the motion model which guided the target.
It can be observed that the the actual motion model under which the
target was simulated to move and the estimated model are the same in most time steps.

\textit{Nonlinear Example}: In this case the motion models and the
measurement models are non-linear, and the unscented Kalman Filter (UKF) 
\cite{Ristic04,UKF} is used for predicting and updating each Gaussian
component in the mixture forward.

The motion models under which the targets are moving are the constant
velocity model and the coordinated turn model with unknown turn rate. The
birth locations are given by $\mathcal{N}( m_4, P_{NL}), \mathcal{N}( m_5,
P_{NL}), \mathcal{N}( m_6, P_{NL})$ where, 
\begin{equation*}
m_4= [ 40000, 0, -50000, 0, 0], \:\: m_5 = [ -50000, 0, 40000, 0, 0],
\end{equation*}
\begin{equation*}
m_6\!=\![\!-10000,\!0,\!0,\!0,\!0],\!P_{NL}\!\!=\!diag([\!1000,\!300,\!1000,%
\!300,\!-1 \times 10^{-4}]).
\end{equation*}
The state vector includes the turn rate in addition to the positions and
velocities in $x,y$ directions and $Q_{NL}$ is the process noise co-variance
matrix.

The observation region is the same as in the linear example. The
measurements are obtained using a bearing and range sensor at (0,0) position 
Clutter is poisson distributed uniformly with an average value of 60. The
measurement noise covariance is $diag([\sigma_\theta^2, \sigma_r^2]$ with $%
\sigma_\theta = \pi/180 rads^{-1}$ and $\sigma_r = 20m$. The markovian
transition matrix is given in (\ref{mvt2}).

\begin{equation}  \label{mvt2}
\vartheta(r^{\prime }|r) = M( r, r^{\prime }) \text{ where } M = 
\begin{bmatrix}
0.8 & 0.2 \\ 
0.2 & 0.8%
\end{bmatrix}%
\end{equation}

\begin{equation}  \label{procnoise}
Q_{NL} = \sigma_{vr}^2 
\begin{bmatrix}
T^4/4 & T^3/2 & 0 & 0 & 0 \\ 
T^3/2 & T^2 & 0 & 0 & 0 \\ 
0 & 0 & T^4/4 & T^3/2 & 0 \\ 
0 & 0 & T^3/2 & T^2 & 0 \\ 
0 & 0 & 0 & 0 & T^2%
\end{bmatrix}%
\end{equation}

\begin{figure}[tbp]
\includegraphics[scale=0.23, trim= 120 50 80 100]{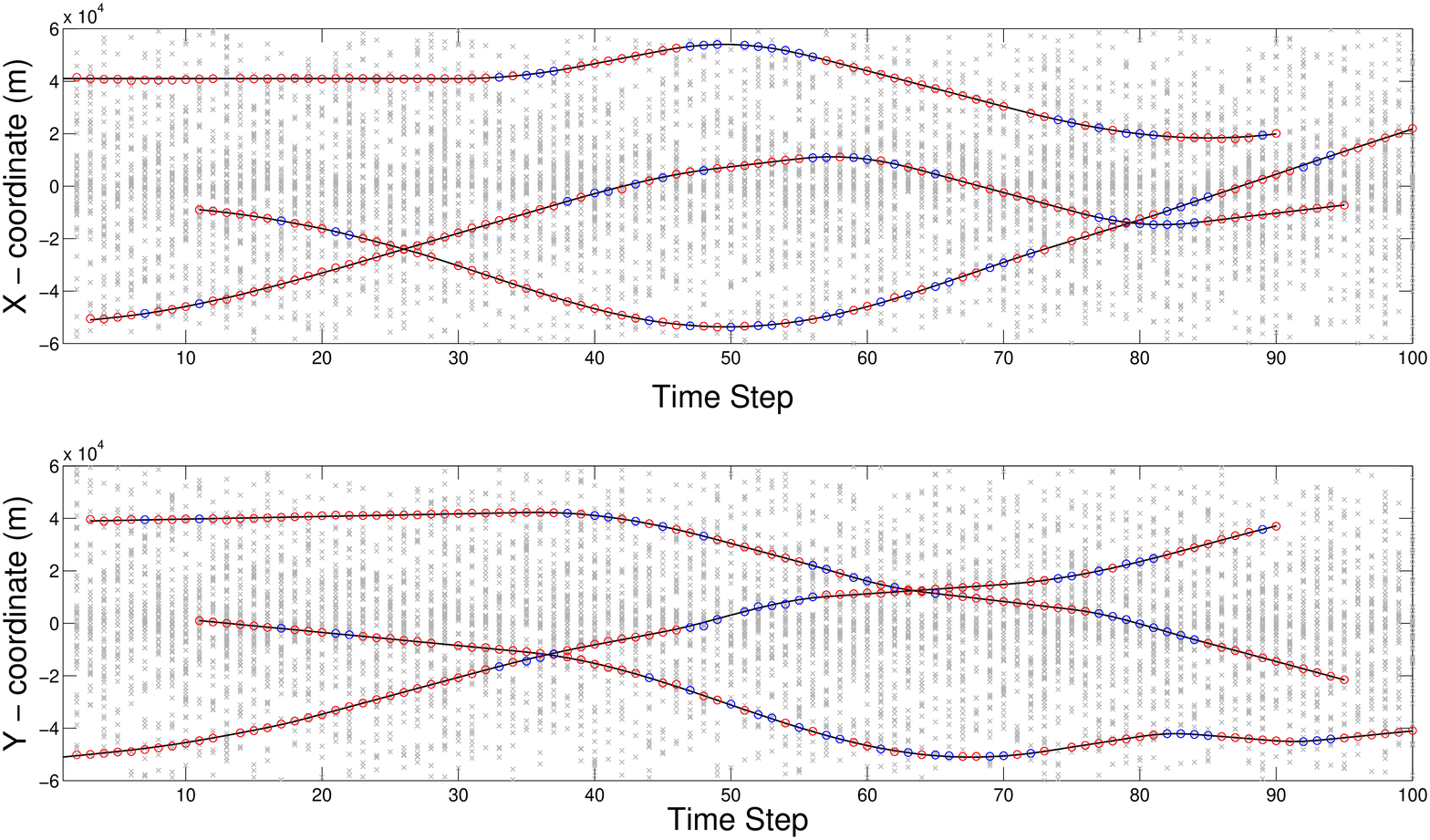}
\caption{ Position Estimates - Non Linear Example.}
\label{est2}
\end{figure}

The Optimal Subpattern Assignment Metric (OSPA)\cite{OSPApaper} values
calculated for 100 monte carlo runs for the non linear
example are shown in the bottom graph of fig.(\ref{ospa}). The bottom graph of figure (\ref{tgt1}) shows the
probabilities of estimating each motion model (colour coded) in each time
step for target 1 in the non-linear
example. It can be observed that the the actual motion model under which the
target was simulated to move has the higher probability.

\section{Conclusion}

An algorithm for tracking multiple maneuvering targets is proposed using the
GLMB multi-target tracking filtering with JMS motion models. Analytic
prediction and update equations are derived along with Linear Gaussian and
Unscented implementations. Simulation results verify accurate tracking and
motion model estimation.

\end{document}